\documentclass[%
reprint,
twocolumn,
amsmath,amssymb,
aip,pop,
nofootinbib,
longbibliography, 
floatfix,
]{revtex4-2}
\usepackage[utf8]{inputenc}
\usepackage{lipsum}
\usepackage{graphicx}
\usepackage{dcolumn}
\usepackage{bm}
\usepackage{amssymb}
\usepackage{amsmath}
\usepackage{textcomp}
\usepackage{color}
\usepackage{mathtools}
\usepackage{units}
\usepackage{appendix}

\begin{document}

\title{On parallel laser beam merger in plasmas}

\author{K. V. Lezhnin}
\email{klezhnin@pppl.gov}
\affiliation{Princeton Plasma Physics Laboratory, 100 Stellarator Rd, Princeton, NJ 08540, USA}
\author{Kenan Qu}
\affiliation{Department of Astrophysical Sciences, Princeton University, Princeton, NJ 08544, USA}
\author{N. J. Fisch}
\affiliation{Department of Astrophysical Sciences, Princeton University, Princeton, NJ 08544, USA}
\author{S. V. Bulanov}
\affiliation{ELI Beamlines Facility, The Extreme Light Infrastructure ERIC, Za Radnicí 835, Dolní Břežany 25241, Czech Republic}
\affiliation{National Institutes for Quantum and Radiological Science and Technology (QST), Kansai Photon Science Institute, Kyoto 619-0215, Japan}

\date{\today}

\begin{abstract}
Self-focusing instability is a well-known phenomenon of nonlinear optics, which is of great importance in the field of laser-plasma interactions. Self-focusing instability leads to beam focusing and, consequently, breakup into multiple laser filaments. The majority of applications tend to avoid the laser filamentation regime due to its detrimental role on laser spot profile and peak intensity. In our work, using nonlinear Schr\"{o}dinger equation solver and particle-in-cell simulations, we address the problem of interaction of multiple parallel beams in plasmas. We consider both non-relativistic and moderately relativistic regimes, and demonstrate how the physics of parallel beam interaction transitions from the familiar self- and mutual-focusing instabilities in the non-relativistic regime to moderately relativistic regime, where an analytical description of filament interaction is impenetrable.

\end{abstract}

\maketitle

\section{Introduction}\label{sec:intro}
Self-focusing or self-modulational instabilities are well-known processes in optical fibers\cite{Agrawal2006} and laser-plasma interaction\cite{Mourou2006}. These instabilities typically lead to the broadening of the pulse spectrum and may cause a breakup of the incident pulse into multiple longitudinal and transverse filaments \cite{Esarey1997}. Such behavior is usually found detrimental to potential applications, and a multitude of approaches to avoid such instabilities are proposed. For instance, in works by Kalmykov et al.\cite{Kalmykov2008}, it is shown that the introduction of an auxiliary parallel laser pulse with a shorter wavelength may help to control plasma wave beating and to balance out the self-focusing effect to facilitate steady propagation of the laser beam without significant changes in laser spot over multiple Rayleigh lengths, potentially improving the laser wakefield acceleration of electrons. In some applications, it is possible to use incoherent beams \cite{Kato1984,Edwards2017}, thereby effectively avoiding the power threshold for the self-focusing instability, $P_{\rm cr}=17.3 {\rm GW} \cdot (n_{e}/n_{\rm cr})^{-1}$ ($n_e$ is the plasma density and $n_{\rm cr}=m_e\omega_0^2/4\pi e^2$ is the critical density for the laser pulse of frequency $\omega_0$), see Refs. \cite{Agrawal2006,Mourou2006,Esarey1997}. Oblique crossing of multiple laser pulses for the self-focusing suppression is also discussed theoretically \cite{MalkinFisch2020}. As the majority of potential applications benefit from higher laser powers/intensities on target, including laser ion acceleration, gamma ray and $e^--e^+$ pair generation, inertial confinement fusion, and plasma-based laser amplification, it is very instrumental to control the self-focusing instability.

Coming from a different perspective, it is well-known that one may use a mutual-focusing-like instability to avoid catastrophic self-focusing in air by redistributing the laser power into so-called parallel beam arrays that merge in a controllable fashion, see Ref.~\cite{Fairchild2017,Reyes2020} and references therein. For instance, in Ref.~\cite{Fairchild2017}, it was shown that such beam combination is experimentally feasible for two subcritical laser filaments ($P_1,P_2<P_{\rm cr}$) with total power $P_1+P_2$ exceeding $P_{\rm cr}$, and, thus, triggering mutual focusing and eventual beam merger. Such phenomenon was explained theoretically using nonlinear Schr\"{o}dinger equation (NSE) envelope model. In Refs.~\cite{McKinstrie1988,Berge1998}, a merger of collinear beams via self-modulational instability was discussed for a Kerr-like medium, while Ref.~\cite{Berge1997} addresses the question of an interaction of the parallel beams with a transverse shift. One may expect that similar physics will prevail in tenuous plasmas as well, as long as $a_0 \ll 1$ ($a_0$ is the dimensionless laser field, $a_0 \equiv eE_0/m_e\omega_0c$, subscript "0" denotes the initial laser field value). When laser amplitude becomes $a_0 \approx 1$, the approximation of the nonlinear Schr\"{o}dinger model with cubic nonlinearity breaks up, as the self-focusing term, $|a|^2$, is assumed to be much smaller than one. The theoretical description of the laser beam of high intensity propagating in tenuous plasmas thus becomes incomprehensible for the NSE model, and kinetic simulations are usually utilized to consider such a regime. Pioneering works by Askaryan et al.~\cite{Askaryan1994,Askaryan1997} showed that an incident beam of $a_0=5$ propagating in near-critical plasma density of $n_e/n_{\rm cr}=0.5625$, while initially being separated into two via filamentation, later on combined two filaments into a single tightly focused beam with negligible power losses. Similar behavior was observed in 3D PIC simulations with $a_0\sim 1$\cite{Naumova2002} and $a_0\gg 1$\cite{Gu2021}. Thus, it may be possible to find a regime to reliably combine multiple parallel beams in collisionless plasmas to achieve higher pulse powers and avoid energy losses associated with the laser power transmission through plasmas.

In this paper, we address the problem of parallel beam combination in the regime of $a_0 \lessapprox 1$, i.e. in the regime where the NSE model may be applied to an extent, but ultimately it fails as soon as field amplitude reaches $a_0 \approx 1$. First, we recall the criteria for parallel beam merger in 2D (one transverse dimension) and 3D (two transverse dimensions), reproducing or closely following Ref.~\cite{Berge1997}. Then, by using the NSE solver, we find a threshold of beam combination numerically, while also analyzing the role of the ponderomotive effect on the beam combination. Next, we run two- and three-dimensional fully kinetic relativistic Particle-In-Cell (PIC) simulations and demonstrate the mechanism of beam merger, and show the scalability of the process to higher laser powers (including overcritical) and relativistic intensities ($a_0>1$). The beneficial role of beam combination for the suppression of the power propagation losses is also highlighted.

This paper is organized as follows. We start by recalling critical relativistic self-focusing power, restating the criterion of beam combination in 3D and deriving the same properties for 2D in Section~\ref{sec:theory}. Next, Section~\ref{sec:NSEresults} is devoted to the discussion of NSE simulations, which help to find the critical beam separation and check the importance of the ponderomotive effect for beam combination. In Section~\ref{sec:PICresults}, we discuss the results of 2D/3D PIC modeling of parallel beam interactions in both $a_0<1$ and $a_0>1$ regimes. We conclude by discussing the limitations of the beam combination approach and the path towards the experimental investigation of the aforementioned phenomena in Section \ref{sec:diss}.

\section{Theoretical background: NSE model and threshold for beam combination}
\label{sec:theory}

Let us start by recalling one of the basic properties of the nonlinear Schr\"{o}dinger equation (NSE) model, namely, critical power for self-focusing. We will consider the NSE in the following form:

\begin{equation}
    \left[\frac{\partial}{\partial t} + v_g \cdot \nabla - \frac{ic^2}{2\omega}\nabla^2_\perp - \frac{ic^2 \omega_{pe}^2}{2\omega^3}(\hat{v_g} \cdot \nabla)^2 - \frac{i \omega_{\rm pe}^2}{8 \omega}|a|^2 \right] a = 0.
\end{equation}

\noindent Here, $a$ is the laser field envelope (electric field normalized to $m_e \omega c/e$), $\omega$ is the laser frequency, $\omega_{\rm pe}^2 = 4\pi n_e e^2/m_e$ is the squared plasma frequency, $v_g$ and $\hat{v_g}$ are group velocity vector and unit vector along the laser group velocity, respectively. The first and the second terms correspond to the envelope propagation, third - diffraction, fourth - group velocity dispersion (GVD), and fifth - self- and mutual-focusing term.

In what follows, we normalize spatial coordinates to $c/\omega$, temporal - to $\omega^{-1}$, and shift to the reference frame moving with $v_g$. We also consider both 1D+1T (one transverse spatial dimension and one temporal dimension) and 2D+1T NSE models, i.e., we solve for the evolution of the beam cross-section, as well as for the coevolution of the beam longitudinal and transverse profiles. In the case of the 1D+1T model, it yields:

\begin{equation}
    \left[\frac{\partial}{\partial \tau} - \frac{i}{2}\frac{\partial^2}{\partial y^2} - \frac{i \omega_{pe}^2}{2\omega^2}\left[ 1-\frac{\omega_{\rm pe}^2}{\omega^2}\right]\frac{\partial^2}{\partial x^2} - \frac{i \omega_{\rm pe}^2}{8 \omega^2}|a|^2 \right] a = 0,
    \label{eqn:NSE1D1T}
\end{equation}

\noindent with the $x$-axis being the laser axis and the $y$-axis being the transverse axis. This equation is further solved numerically in the current form and in an extended model involving density perturbation. This model effectively corresponds to 2D geometry, which will also be considered in 2D PIC simulations.

The 2D+1T model looks as follows:

\begin{equation}
    \left[\frac{\partial}{\partial \tau} - \frac{i}{2}\nabla^2_\perp  - \frac{i \omega_{\rm pe}^2}{8 \omega^2}|a|^2 \right] a = 0.
    \label{eqn:NSE2D1T}
\end{equation}

\noindent This model is relevant for the 3D geometry and will be addressed theoretically and numerically with the NSE solver.

Let us recall the method to derive the self-focusing threshold, as it is also used to derive the beam combination threshold in 2D and 3D. Following Ref.~\cite{SulemSulemNSE}, we first write the Hamiltonian corresponding to the Equation~\ref{eqn:NSE2D1T} (we further denote $\alpha \equiv \omega_{\rm pe}^2/8\omega^2$):

\begin{equation}
    H =\frac{1}{2} \int \left[|\nabla a|^2 - \alpha |a|^4 \right] d{\bf x},
\end{equation}

\noindent where the integration is performed over one or two transverse dimensions in the case of 1D+1T and 2D+1T models, respectively. Using the variance identity, $V$, for the envelope $a$ (see Chapter 2.4 in Ref.~\cite{SulemSulemNSE}), we could write down the self-focusing/combination criterion as follows:

\begin{equation}
    \frac{d^2V}{dt^2} = 8H - 2 \alpha(d-2) \int |a|^4 d{\bf x}=0.
    \label{eqn:variance}
\end{equation}

\noindent Here, $d$ is the number of transverse dimensions. For 2D+1T, d=2, and, assuming the Gaussian profile of the laser electric field, $a = a_0 \exp{[-r^2/w^2]}$, it leads to the threshold for beam self-focusing:

\begin{equation}
    a_0^2 w^2 = 4/\alpha,
\end{equation}

\noindent which could be written in terms of critical power in dimensional units as a well known-result:

\begin{equation}
    P_{\rm cr} = \frac{2 m_ec^3}{r_e}\frac{n_{\rm cr}}{n_e} = 17.5 {\rm GW} \cdot \frac{n_{\rm cr}}{n_e}.
\end{equation}

\noindent Here, $r_e = e^2/m_ec^2$ is classical electron radius.

Interestingly, a similar approach is applicable to find the combination threshold of two shifted envelopes \cite{Berge1997}. For two envelopes given by $a_{1,2} = a_0 \exp{[-({\bf r}\mp \mathbf{r_c})^2/w^2]}$, with $\pm \mathbf{r_c}$ being the center of mass of the particular envelope and $|\mathbf{r_c}| = \delta$, one could get the following implicit expression for the critical beam separation, $\delta/w \equiv t$:

\begin{align} 
    &\frac{P}{P_{\rm cr}}-1 = \exp{[-2t^2]}(1-2t^2)-3\frac{P}{P_{\rm cr}}\exp{[-4t^2]} \nonumber \\
    &-4\frac{P}{P_{\rm cr}}\exp{[-3t^2]}.
    \label{eqn:mergercond3D}
\end{align}

\noindent Here, $P$ denotes the laser power of a single laser filament. This equation has real positive solutions only for $P>P_{\rm cr}/4$, meaning that pulses below that value do not combine and just diffract around their respective centers of mass; at the same time, for $P>P_{\rm cr}$, $\delta/w \rightarrow \infty$, meaning that two beams do not combine and experience independent self-focusing. These results identically reproduce the results by Ref.~\cite{Berge1997}.

Similar results can be obtained for the 1D+1T model. Taking d=1 in Eqn.~\ref{eqn:variance} and considering envelope $a = a_0 \exp{[-y^2/w^2]}$, one gets the following critical condition for self-focusing:

\begin{equation}
    a_0^2 w^2 = 2\sqrt{2}/\alpha.
\end{equation}

Now, applying Eqn.~\ref{eqn:variance} to the sum of two shifted envelopes in 1D ($a_1 = a_0 \exp{[-(y-\delta)^2/w^2]}$, $a_2 = a_0 \exp{[-(y+\delta)^2/w^2]}$, $a=a_1+a_2$), we get a very similar implicit expression for the threshold beam separation in 2D:

\begin{align} 
    &\frac{P_{2D}}{P_{\rm cr,2D}}-1 = \exp{[-2t^2]}(1-4t^2)-3\frac{P_{2D}}{P_{\rm cr,2D}}\exp{[-4t^2]} \nonumber \\
    &-4\frac{P_{2D}}{P_{\rm cr,2D}}\exp{[-3t^2]}.
    \label{eqn:mergercond2D}
\end{align} 

\noindent Here, $P_{2D}/P_{\rm cr,2D} \equiv a_0^2 w^2/(2\sqrt{2}/\alpha)$. Solving for $\delta/w$ as a function of $P_{2D}/P_{\rm cr,2D}$, we get similar thresholds at $P_{2D}/P_{\rm cr,2D} =1/4$ and $1$.

\section{NSE simulations of beam merger}
\label{sec:NSEresults}

To illustrate the physics of the parallel beam merger, we conduct NSE simulations using a symmetrized split-step Fourier approach \cite{Agrawal2006} implemented in a Python solver \cite{Lezhnin2023}. We address three models: 2D+1T with no contribution from ponderomotive force by solving Eqn.~\ref{eqn:NSE2D1T}, 1D+1T with no contribution from ponderomotive force by solving Eqn.~\ref{eqn:NSE1D1T}, and 1D+1T with the contribution from ponderomotive force \cite{Sprangle2001}, solving:

\begin{eqnarray}
    \left[ \frac{\partial}{\partial \tau}+\frac{v_g}{c}\frac{\partial}{\partial x}- \frac{i}{2 }\frac{\partial^2}{\partial y^2} - \frac{i \omega_{\rm pe}^2}{8\omega^2} (|a|^2 -4 \delta n) \right] a = 0, \label{eqn:nsepdmtv1} \\
    \frac{\partial^2}{\partial \tau^2}\delta n + \frac{\omega_{\rm pe}^2}{\omega^2} \delta n = \frac{1}{4} \nabla^2 |a|^2,
    \label{eqn:nsepdmtv2}
\end{eqnarray}

\noindent which is also coupled with the initial condition of two Gaussian beams with transverse shift. Here, we normalize time to $\omega^{-1}$, spatial coordinate to $c/\omega$, electric field is normalized to $m_e \omega c/e$, density perturbation $\delta n$ - to initial plasma density $n_0$. For simplicity, we assume $\lambda =0.8 \rm um$ and use dimensional units, which should be directly comparable with PIC results presented later on in the manuscript.

To determine the threshold beam separation as a function of laser power, we conduct a scan over dimensionless envelope amplitudes, $a$, from 0.1 to 0.3, and beam half-separation, $\delta$, from 0.5 to 2 beam widths. Pulse width is specified to be equal to 20 um, pulse duration - 5.396 um (30 fs). We initialize two pulses with some beam separation and fixed total energy being equal to the energy of a single pulse with the aforementioned parameters. We specify the grid of 256 by 256 grid nodes and 200 by 200 unit lengths, choose timestep to be equal to $c\cdot dt=dx=0.78$ um, apply periodic boundary conditions, and we also shift to the simulation window moving with $v=v_{g}$.

\begin{figure}
    \centering
    \includegraphics[width=\linewidth]{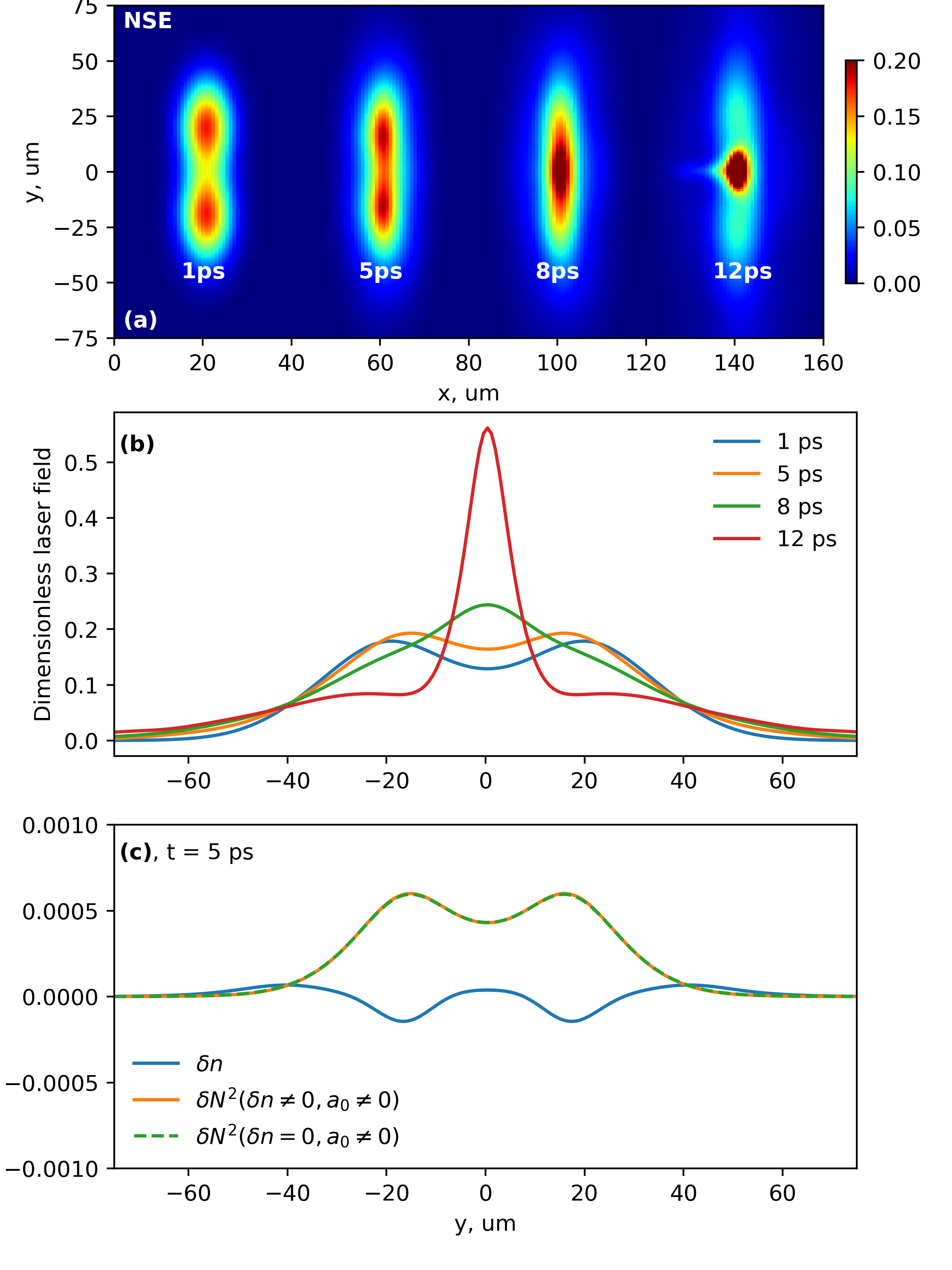}
    \caption{NSE run with $a_0 =0.17$ and beam separation $\delta = w$. Gradual beam merger is seen. (a) 2D envelopes, (b) 1D cuts of NSE envelopes, and (c) the relative role of density and gamma factor in refractive index perturbations are shown.}
    \label{fig:NSEsnapshot}
\end{figure}

Figure~\ref{fig:NSEsnapshot} presents evolution of two beam envelopes with $a_0=0.17$, beam width $w=20$ um, and beam half-separation $\delta=w$, under the Eqns.~\ref{eqn:nsepdmtv1}-\ref{eqn:nsepdmtv2}. One may see how the envelopes collapse into one under the effect of the nonlinear term. As is common in the self-focusing and mutual focusing instabilities, the focusing effect may be understood in terms of modulations of the refractive index. Square of unperturbed plasma refractive index, $N^2_0 = 1 - n_{e0}/n_{\rm cr}$, where $n_{\rm cr} = m_e\omega^2/4\pi e^2$ and $n_{e0}$ is electron number density of unperturbed electron plasma. A perturbed value of the refractive index in the laser field may be written as $N^2 = 1 - n_{e}/\langle \gamma_e \rangle n_{\rm cr}$, which includes both perturbations in density ($n_e = n_{e0}+\delta n$) and mean electron gamma factor, which is connected to laser field by $\langle \gamma_e \rangle = \sqrt{1+a_0^2}$. Figure~\ref{fig:NSEsnapshot}c highlights the relative role of density and gamma factor perturbations at t=5 ps. One may see the importance of the gamma factor contribution and a relatively minor role of density perturbations. Also, one may see the positive density perturbation between two pulses, which translates into negative refractive index perturbation, slightly impeding the pulse combination.

\begin{figure}
    \centering
    \includegraphics[width=\linewidth]{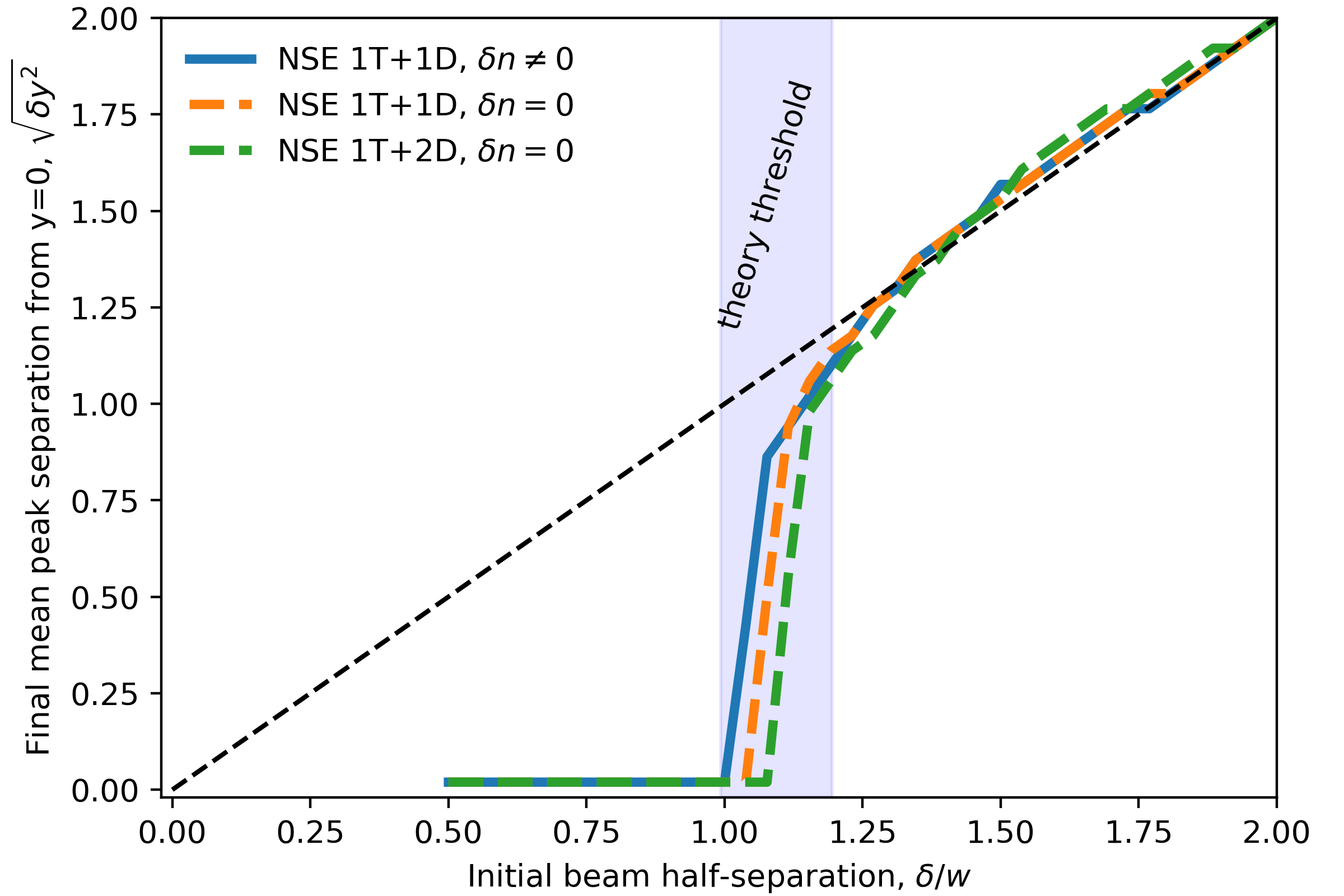}
    \caption{NSE scan on pulse separation for $a_0 =0.175$ (1D+1T) and $a_0=0.209$ (2D+1T) for three types of NSE models. Final beam separation is shown for t=6 ps. The transition from self-focusing to mutual focusing is seen around $\delta\approx w$. The theoretical beam combination threshold (shaded blue region) calculated from Eqns.~\ref{eqn:mergercond3D} and~\ref{eqn:mergercond2D} is specified.}
    \label{fig:NSEscan1D}
\end{figure}

Fixing total laser power and scanning over the beam half-separation $\delta$ using three aforementioned models, we generate Figure~\ref{fig:NSEscan1D}, which aims at seeking the threshold beam separation to still merge two laser beams into one. We find that the threshold half-separation is around $w$. Density perturbations are seen to counteract beam merger, leading to slightly smaller threshold beam separation than in the model with $\delta n=0$. Considering the cross-section of the two-pulse system, we see that the threshold is slightly larger than $w$. Threshold beam separation obtained from NSE scans is in fair agreement with theoretical estimates calculated from Eqns.~\ref{eqn:mergercond3D} and ~\ref{eqn:mergercond2D} (shaded blue region in Fig.~\ref{fig:NSEscan1D}).

\begin{figure}
    \centering
    \includegraphics[width=\linewidth]{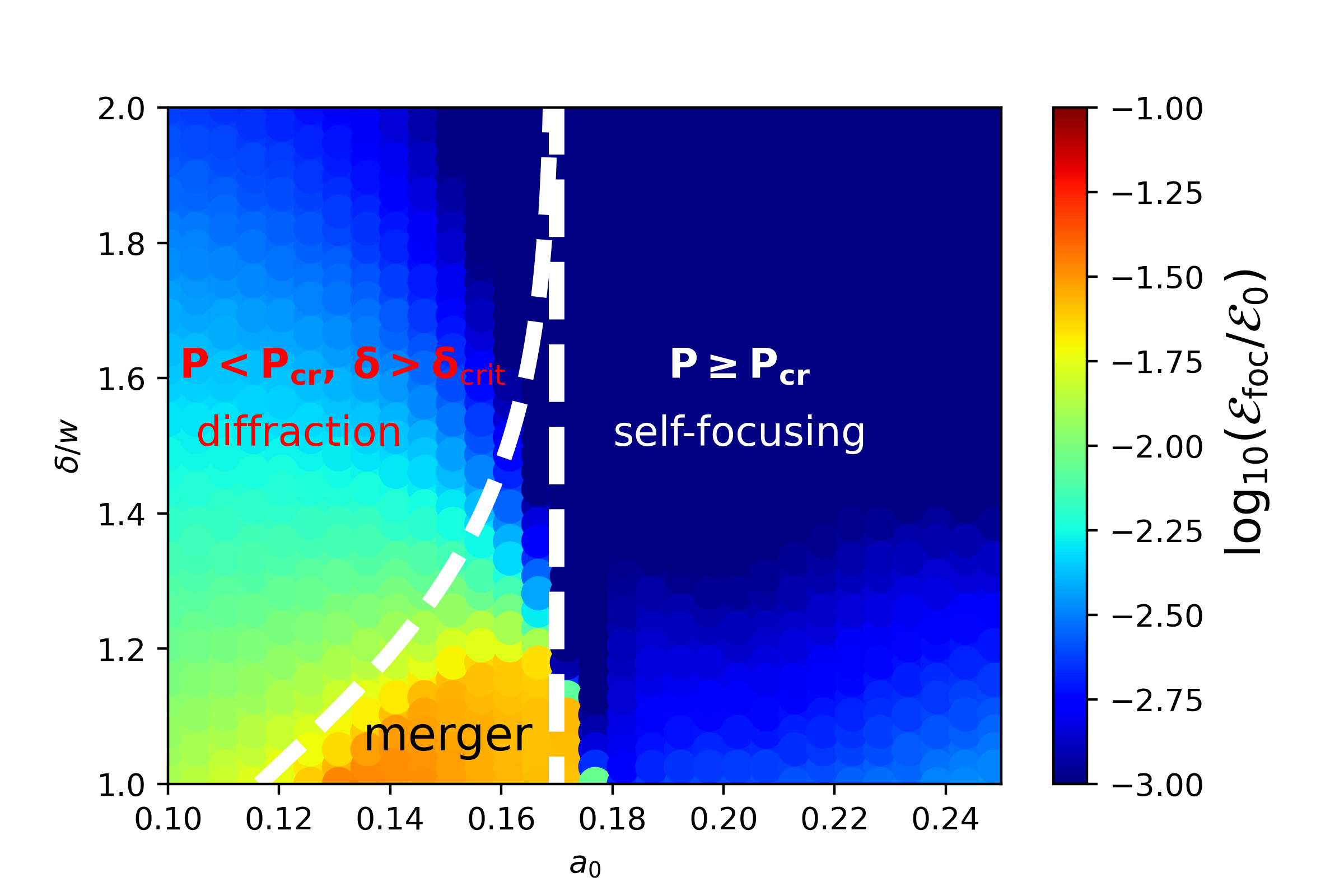}
    \caption{NSE scan on pulse separation, $\delta/w$, and laser field, $a_0$. White dashed lines draw a critical power threshold, $P_{2D} = P_{\rm crit,2D}$, and critical beam separation given by Eqn.~\ref{eqn:mergercond2D}, $\delta = \delta_{\rm crit}$. Regions with diffraction, self-focusing, and beam merger being dominant, are specified.}
    \label{fig:NSEscan2D}
\end{figure}

To better represent the outcomes of the two-pulse interaction, we conduct a 2D scan of the beam field and separation using the NSE model with density perturbations. Figure~\ref{fig:NSEscan2D} represents the results of such scan, with $x$ axis being dimensionless laser field $a_0$, $y$ axis - initial beam half-separation normalized to beam width, $\delta/w$, and color depicts the combination metric by estimating the amount of total beam energy focused to the center of the simulation box. One may see that there is a transition between the regimes with individual beam self-focusing (to the right from $P_{\rm 2D}=P_{\rm cr,2D}$ line), beam diffraction ($\delta > \delta_{\rm crit}$), and beam merger ($\delta \leq \delta_{\rm crit}$, $P_{\rm 2D}<P_{\rm cr,2D}$), which are separated by white dashed lines. These lines are given by $P_{2D}/P_{\rm crit,2D}=1$ and Eqn.~\ref{eqn:mergercond2D}. It implies that by specifying laser pulses with $P\leq P_{\rm cr}$ with separation $\delta\leq \delta_{\rm crit}$, we may expect coalescence of these pulses in one. It should be noted that here we talk about the individual pulse powers. We thus would expect pulse merger as soon as two pulses are close enough ($<1.5 w_0$) and possess total power of $P\sim P_{\rm cr}$, with individual pulses being undercritical.

\section{PIC simulations of beam merger}
\label{sec:PICresults}

{To consider the full complexity of parallel beam interaction, we conduct Particle-In-Cell simulations using the code EPOCH \cite{Arber2015}. The simulation setup is as follows. We shoot two parallel laser beams of $\lambda = 0.8\mu \rm m$, $I= 6\cdot10^{16} \rm W/cm^2$ peak intensity each along the $+x$ axis. Pulse duration is $\tau=30 \rm fs$ (FWHM), and beam waist ($1/e$) is $w=20 \mu \rm m$. This corresponds to $a_0=0.17$ in vacuum and $P/P_{\rm cr,2D}\approx 1$. The beam separation is chosen to be equal to $\delta = w =20 \rm um$. We also conduct analogous runs with a single beam with the same beam width and duration and pulse energy matched to the two-laser case ($I_{\rm aux}=I_1+I_2$) or matched to the energy of one of those pulses ($I_{\rm aux}=I_1=I_2$). The target is the uniform semi-infinite plasma slab with immobile ions, $n_e/n_{\rm cr}=0.032$. The physical parameters are similar for 2D and 3D runs, and only numerical parameters are changed to ensure the reasonable computational cost of a three-dimensional simulation. In 2D, the grid resolution is $20$ grid nodes per micron, box size is $100 \mu \rm m \times 100 \mu \rm m$, number of particles per cell per species is fixed to 2000. We also conducted a convergence study with higher grid resolution 40 grid nodes per micron and 200 particles per cell and smaller particle resolution (20 grid nodes per micron and 20,100,1000 particles per cell) to verify the persistence of physics of the observed refractive index perturbations. In 3D, the longitudinal grid resolution is 12 per micron and 6 per micron in each transverse direction. Number of particles per cell is equal to 4. We adopt a moving window setup, starting to move the simulation window with the group velocity of the laser pulse, $v_g=c\cdot \sqrt{1-n_e/n_{\rm cr}}$, as soon as the laser pulse reaches 2/3 of the simulation box length. Simulation time is 12 picoseconds.}

\begin{figure}
    \centering
    \includegraphics[width=\linewidth]{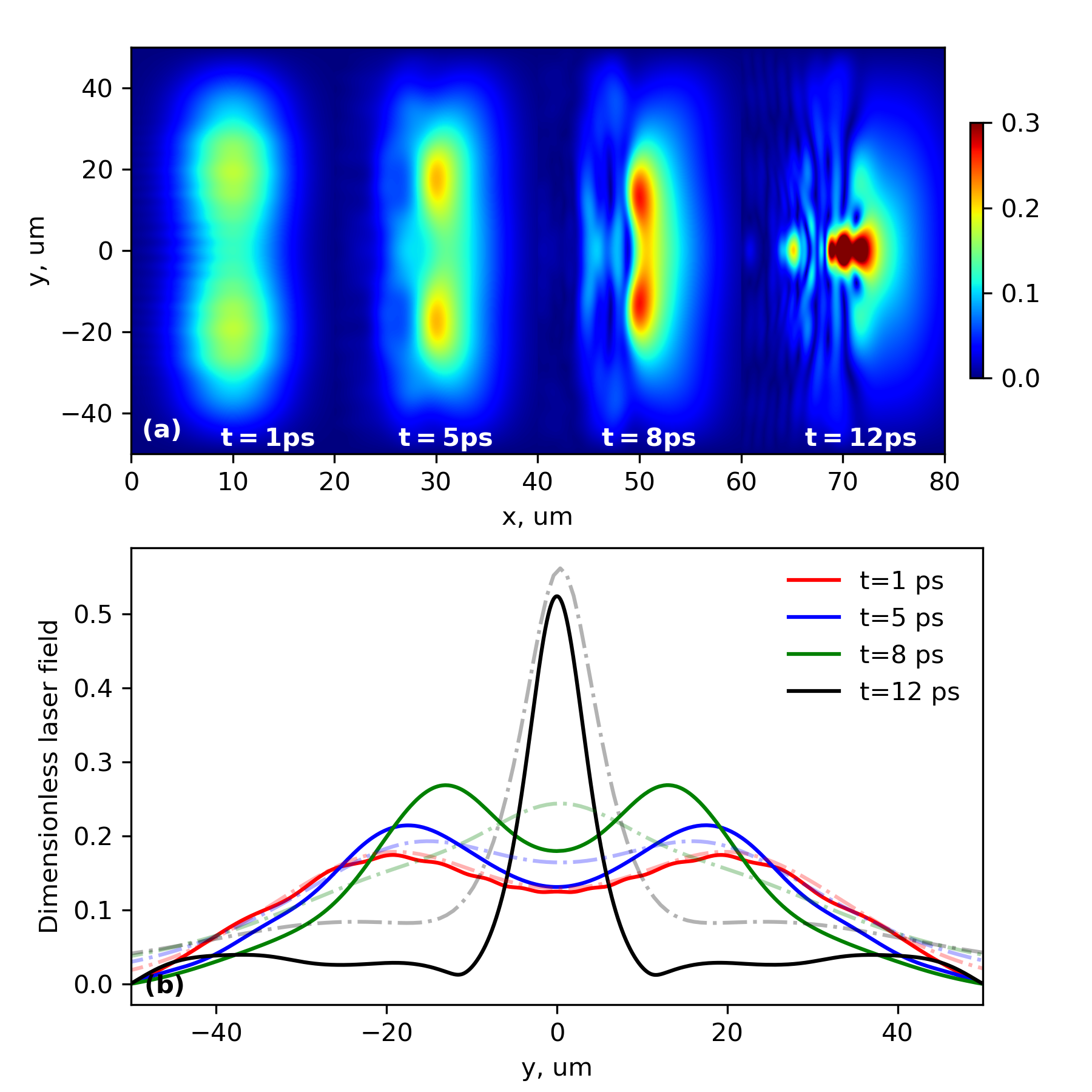}
    \caption{2D PIC simulation of the evolution of two laser envelopes from 1 ps to 12 ps into the run superimposed on a single spatial domain. Self-focusing, beam migration, and beam combination is seen. Solid-dashed lines depict the result of a corresponding NSE simulation.}
    \label{fig:PICsnapshot}
\end{figure}

Let us first consider a 2D simulation with beam half-separation $\delta= 20 \mu {\rm m}= w$. Figure~\ref{fig:PICsnapshot} illustrates the process of coalescence of two parallel laser beams. Here, we see initial laser envelopes at t=1 ps, self-focusing stage at t=5 ps, beam migration at t=8 ps, and full coalescence at t=12 ps. The process and merging timescale are similar to the one in the NSE case, as one may see in Fig.~\ref{fig:NSEsnapshot} and via solid-dotted lines in Fig.~\ref{fig:PICsnapshot}b. Still, the comparison is complicated by at least two factors: (I) during the self-focusing stage, both laser pulses reach dimensionless amplitudes of around 0.4, which formally violates the NSE model assumption of $|\hat{a}|^2\ll 1$ and (II) the process of laser self-focusing is inseparable from Forward Raman Scattering (FRS), which leads to longitudinal modulations of the laser envelope. To suppress the latter and demonstrate a cleaner picture of beam merging, we considered smaller wavelength ($\lambda=0.8 \mu \rm m$) and shorter pulse duration ($\tau=30 \rm fs$) in comparison to our early simulations with $\lambda=1 \mu \rm m$ and $\tau=100 \rm fs$. Recalling the metric on the interplay between FRS and self-focusing \cite{Mori1997}, $\Gamma \equiv P/1{\rm TW} \cdot \tau/1{\rm ps} \cdot (n_e/10^{19}{\rm cm^{-3}})^{5/2} \cdot (\lambda/1 \mu \rm m)^{4}$, we may see that we are able to get from $\Gamma \approx 5.3$ to $\Gamma \approx 0.65$, i.e. we transition from FRS-dominated regime close to the self-focusing-dominant regime. 

To understand the reasons behind the beam migration towards coalescence, we analyzed density, electron energy, and refractive index perturbations around two laser beams. The refractive index is given by:

\begin{equation}
    N^2 = 1 - \frac{n_e}{n_{\rm cr,rel}} =1 - \frac{\omega_{\rm pe}^2}{\langle \gamma_e \rangle \omega_0^2} = 1 - \frac{n_e}{n_{\rm cr}}\frac{1}{1+\frac{\langle \mathcal{E}_{ke}\rangle}{m_ec^2}}.
\end{equation}

\begin{figure*}
    \centering
    \includegraphics[width=0.45\linewidth]{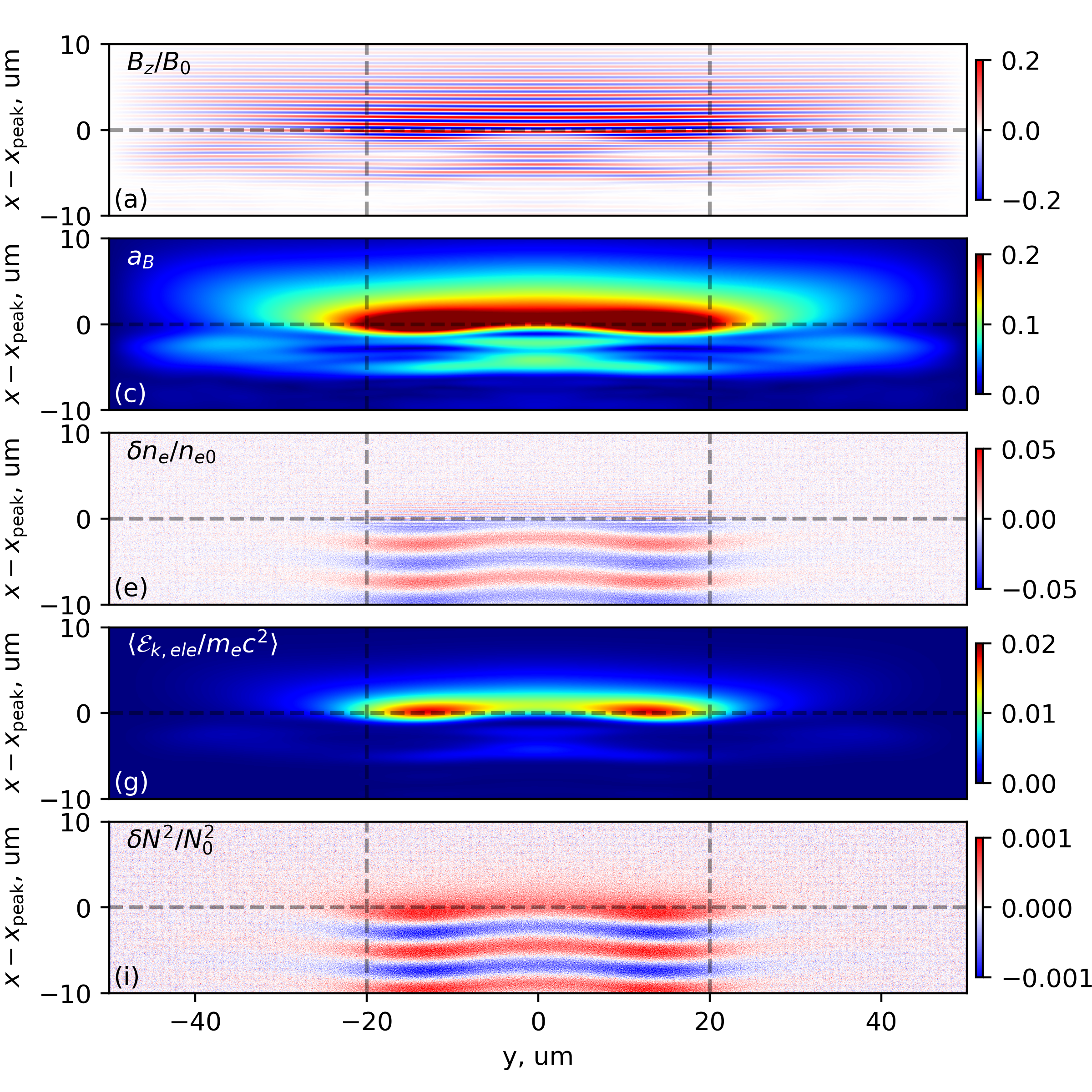}
    \includegraphics[width=0.45\linewidth]{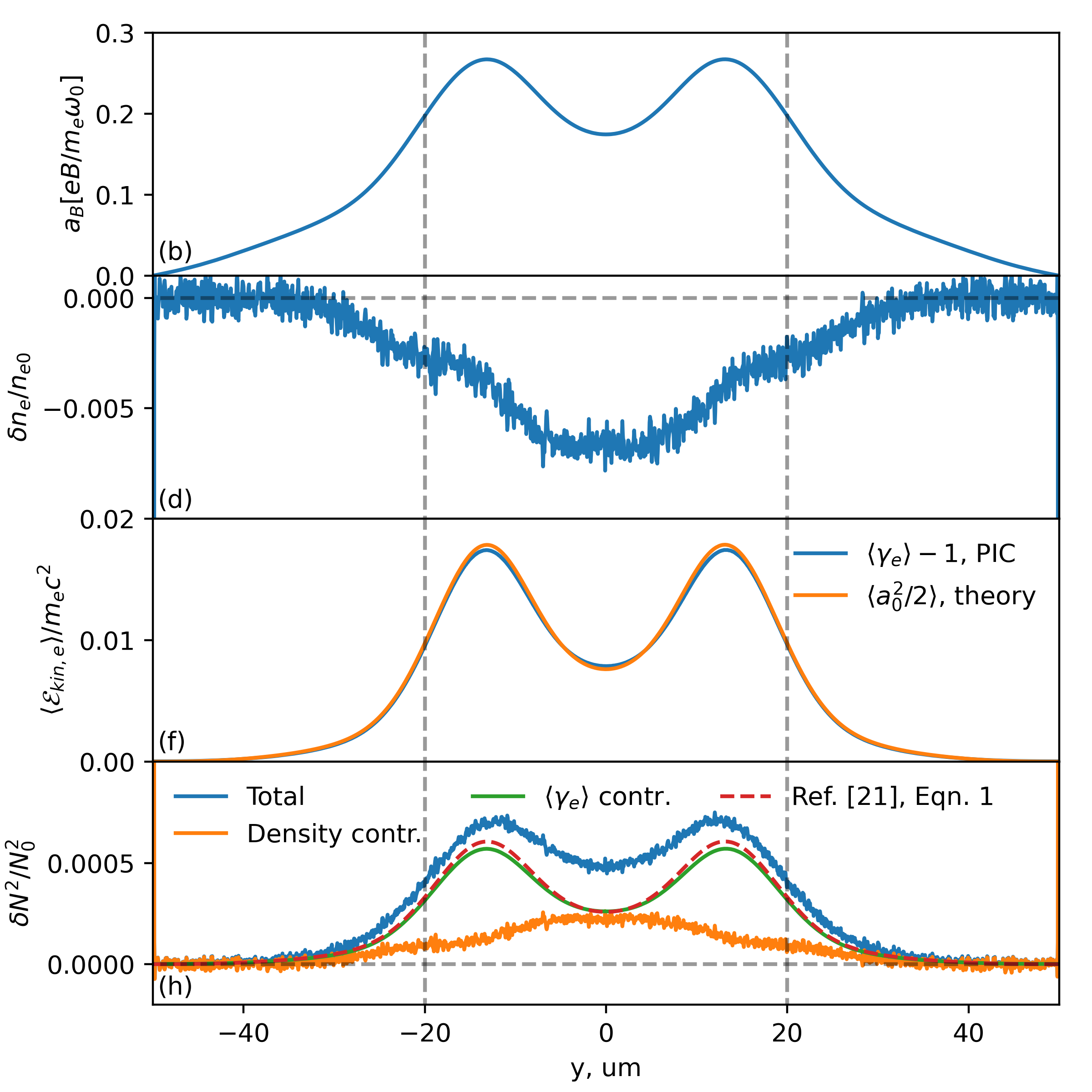}
    \caption{Physics of beam merger in 2D PIC simulations. Beam envelopes, density, electron gamma factor, and refractive index perturbations along with the relative contributions of the respective effects. We also compare the refractive index perturbations with ones predicted theoretically (red dashed curve).}
    \label{fig:PICmerging_mechanism}
\end{figure*}

\noindent Before laser pulses enter the simulation domain, $N^2 \equiv N_0^2 = N^2 (n_e=n_{e0},\langle \gamma_e \rangle \approx 1)=1-0.032=0.968$. Refractive index perturbations are calculated as $\delta N^2/N_0^2 = [N^2(n_e,\langle \gamma_e \rangle) -N_0^2]/N_0^2$. Figure~\ref{fig:PICmerging_mechanism} illustrates laser field (a), laser envelope (b,c), density perturbation (d,e), electron energization (f,g), and refractive index distributions (h,i) in 2D (left) and 1D as cuts at peak laser amplitude along the y axis (right) at t=8 ps. Figs.~\ref{fig:PICmerging_mechanism}a,c depict laser field and envelope, respectively. Dashed vertical lines denote the initial location of two beam envelopes; one may notice that two beams are indeed moving away from their initial laser axes towards amalgamation. It also may be seen from the 1D cut at the peak laser intensity (Fig.~\ref{fig:PICmerging_mechanism}b). Fig.~\ref{fig:PICmerging_mechanism}e shows density perturbation around the pulse envelope, with the density depression at the pulse peaks and plasma wake structure behind the pulses, with the spatial period close to $\lambda_{\rm pe} = \lambda /\sqrt{n_e/n_{\rm cr}}\approx 4.47 \rm um$. One may also notice density perturbation with a spatial period of around $\lambda$, corresponding to electron oscillation in the laser field. The magnitude of perturbation is around $1 \%$, strongly exceeding density perturbations in the NSE case ($<0.1\%$). Fig.~\ref{fig:PICmerging_mechanism}d depicts a 1D density perturbation profile averaged over laser wavelength $\lambda$ along the laser axis. The density dip between laser pulses is notable here, while density depression regions around the laser peaks do not survive the averaging - mainly due to the dominant contribution of strong electron oscillations in the laser field. Fig.~\ref{fig:PICmerging_mechanism}g demonstrates the mean electron kinetic energy profile, and Fig.~\ref{fig:PICmerging_mechanism}f provides a comparison of the electron energization derived from PIC simulation with the theoretical prediction $\langle \gamma_e \rangle \approx 1+\langle a_0^2/2 \rangle$ (both were averaged over $(x_{\rm peak}-\lambda/2,x_{\rm peak}+\lambda/2$)). Decent agreement is seen in all the snapshots from 4 to 9 ps, with stronger deviations appearing once laser amplitude reaches $a_0 \approx 0.3$. Finally, Figs.~\ref{fig:PICmerging_mechanism}h,i show total refractive index perturbation in 2D (i) and 1D cuts of total refractive index perturbation (h, blue solid line), refractive index perturbation due to density perturbation only (h, orange solid line), refractive index perturbation due to electron energization only (h, green solid line), and refractive index perturbation from the theory (h, red dashed line, see Eqn.~1 from Ref.~\cite{Max1974}).

Overall, the structure of the refractive index perturbation and its magnitude are similar to the NSE case, as may be seen by comparing Figure~\ref{fig:PICmerging_mechanism} and Figure~\ref{fig:NSEsnapshot}. In both PIC and NSE models, $\langle \gamma_e \rangle$ contribution to the refractive index modulations is the dominant one. However, electron density contribution is far more noticeable in the PIC case and becomes comparable to $\langle \gamma_e \rangle$ impact at the time of beam amalgamation ($t>9$ ps). Since we are using 2000 particles per cell, with each particle corresponding to 0.05\% of the initial plasma density, we may be confident in the validity of the density profile and its contributions to the refractive index. Throughout the simulation, we observe good agreement between the longitudinally-averaged theory prediction for the refractive index perturbations and the one observed in PIC (also longitudinally-averaged), with a slight increase of refractive index around $y=0$ due to the difference in electron density dynamics. Individual (i.e. non-averaged) profiles may not match, though, partially due to strong density perturbation in the laser field and oscillatory structure of electron energization around the location of the laser peaks. Auxiliary runs with a single pulse with either $I_{\rm aux}=I_1+I_2$ or $I_{\rm aux}=I_1=I_2$ also helped to interpret the beam merger mechanism. Simulation with $I_{\rm aux}=I_1$ is in very good agreement with theory, both in terms of electron heating and density perturbations (and, consequently, refracting index perturbations). On the contrary, $I_{\rm aux}=I_1+I_2= 2 I_1$ simulation is in fair agreement with theory in the initial stages of beam self-focusing but quickly departs due to stronger density perturbations in the PIC model. We may thus conclude that theoretical calculations of average electron energy and refractive index, NSE models, and 2D PIC agree for the laser pulses below $a_{\rm crit}\approx 0.2$, with density perturbations growing significantly for stronger pulses and thus departing from theoretical prediction. As we showed in the presented 2D PIC run, pulses still manage to merge, even though they possessed enough power to self-focus individually, as suggested by theory arguments from Section \ref{sec:theory}. In conclusion, different electron density behavior seems to be the reason for sustained beam merger efficiency for $P/P_{\rm crit,2D}\geq 1$.

\begin{figure}
    \centering
    \includegraphics[width=\linewidth]{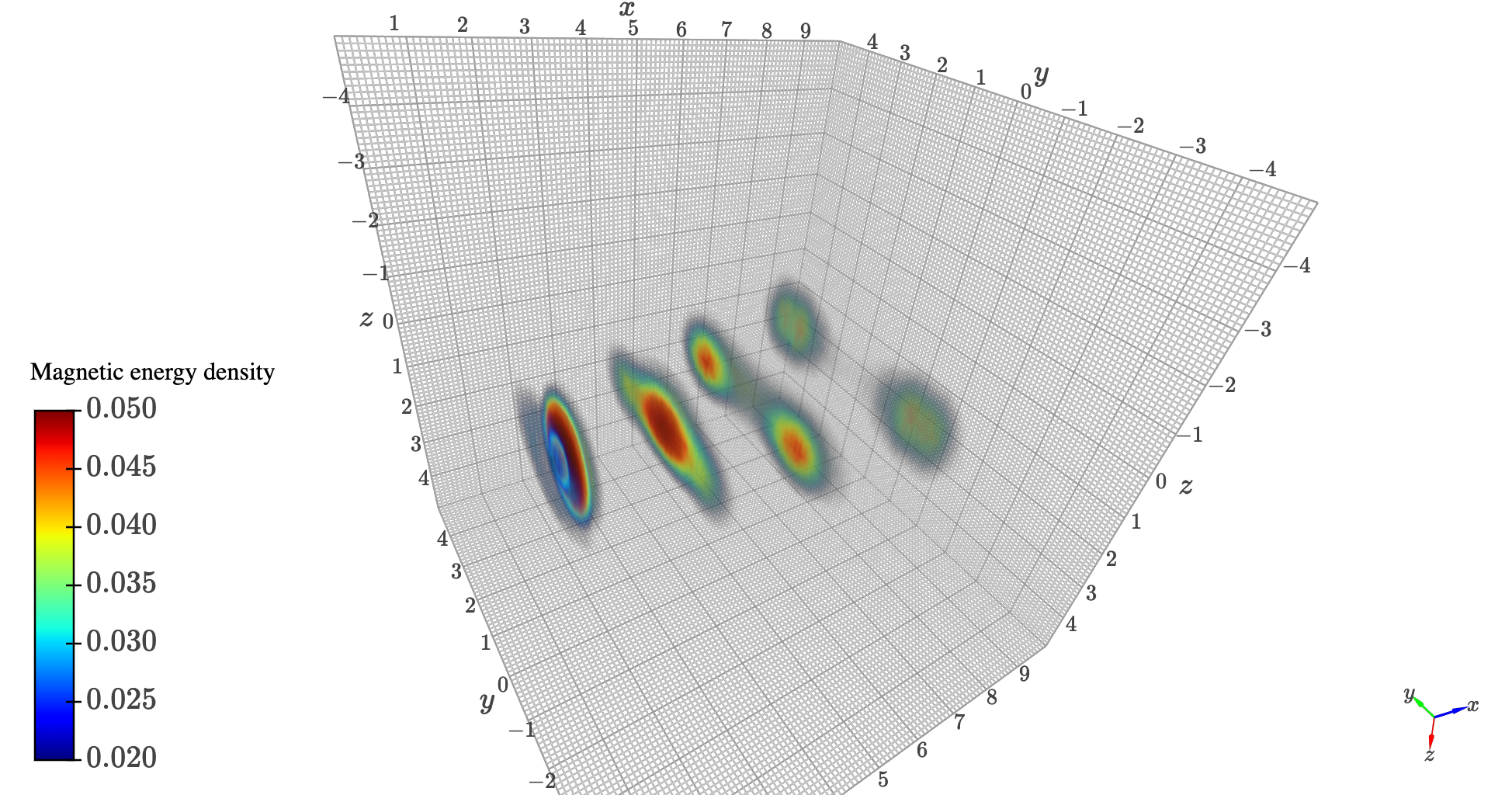}
    \includegraphics[width=\linewidth]{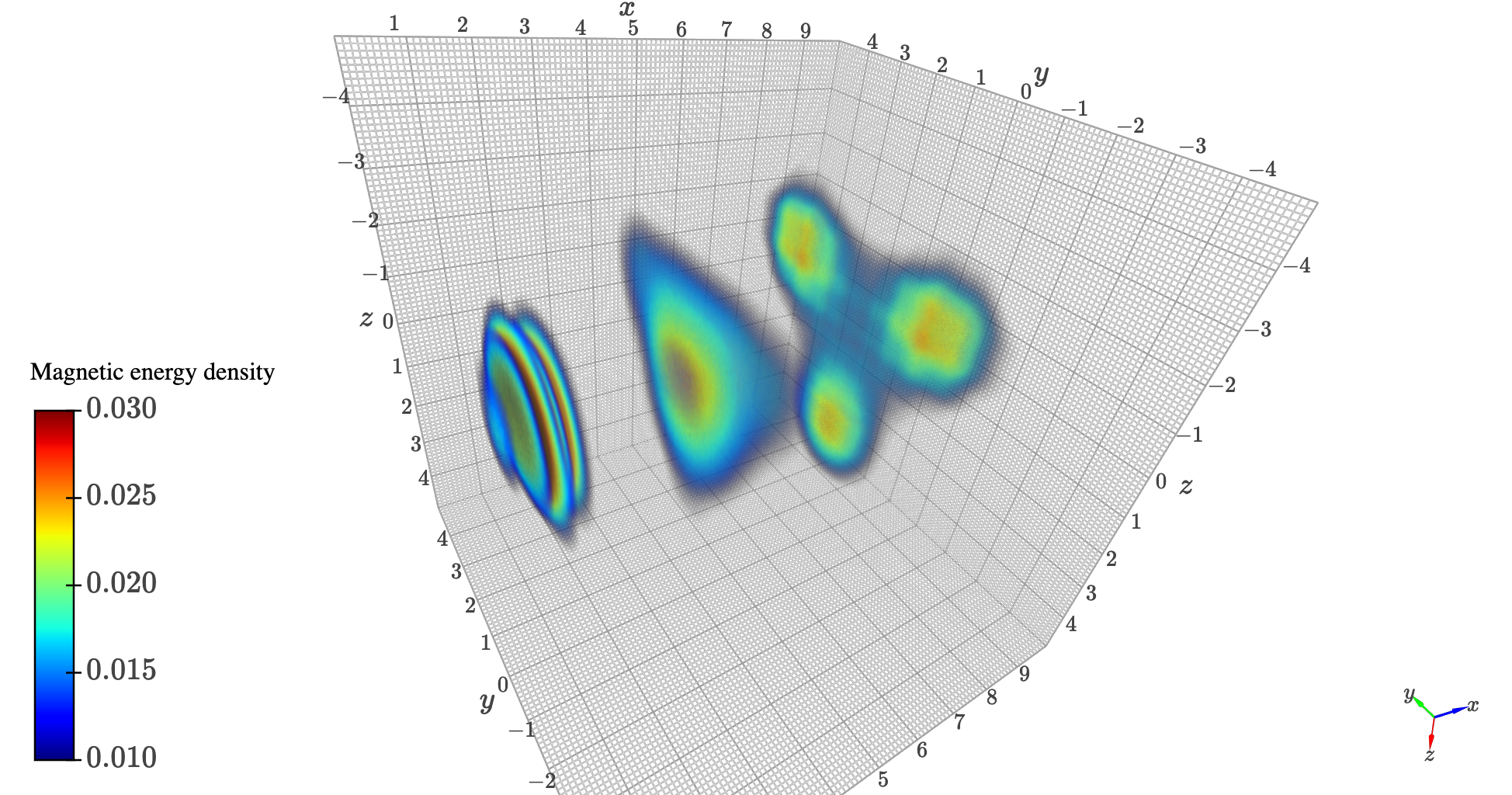}
    \caption{Evolution of magnetic energy density of a system of two/three laser pulses over 10 ps in 3D PIC simulation. Beam propagation, self-focusing, beam migration, and merger are observed.}
    \label{fig:PIC3D}
\end{figure}

A similar pulse merger mechanism is observed in 3D PIC simulation, as one may see in Figure~\ref{fig:PIC3D}a. Here, we show the evolution of the magnetic energy density of two laser pulses over time, from 1 ps (two spots distant from each other and further away from the observer) to 10 ps (single focused beam, appears closest to the observer). As the pulses propagated for $\approx 3$ mm through the plasma, we do not plot the whole box, but rather superimpose the output data from 3D PIC moving window simulations onto the same box of reduced size for clarity. The timescale of the merger of the two-pulse system fairly agrees with the 2D PIC results, which suggests the convergence of the results. Although we do not show the refractive index modulations due to high noise in such diagnostics, we indeed see a similar structure of average electron energization as we identified in the case of 2D PIC (Fig.~\ref{fig:PICmerging_mechanism}). At the same time, due to a small number of particles per cell, the relative role of the density perturbations is indeed overestimated; thus, the 3D run may only be used as an attempt to address the effect of geometry, rather than to understand the details of the refractive index perturbations. As a result of the simulation, we observe the formation of a single beam with the power estimated to be around 1.9 times the power of each input beam. Comparing with the case of a single pulse with the power matching the total power of two pulses, it leads to the stronger development of FRS, which leads to power losses, resulting in the final power (i.e. after propagation through the plasma slab of $\sim 3$ mm at 10 ps into the simulation) of 1.64 times the power of each input beam in the two-pulse case. Thus, by spatially separating two slightly overcritical pulses ($P/P_{\rm cr,3D} \approx 1.4$), we can suppress both FRS and filamentation instability, thus improving the resulting laser pulse power. This is in a way similar to Ref.\cite{Lezhnin2021}, where, by redistributing the total laser power in the frequency domain, we were able to avoid laser power losses due to FRS. Here, we redistribute laser power spatially and combine it back at a given length to have a powerful beam with a reduced amount of power losses. Such a method of avoiding laser power losses was previously used in the works on laser arrays in air, see, e.g., Ref.~\cite{Fairchild2017,Reyes2020}; here, we demonstrated that we can utilize a similar approach for high power lasers propagating in tenuous plasmas. The considered approach may be of use for the inertial confinement fusion experiments, where laser pulse instabilities are known to limit laser power delivery to the target and cause unwanted asymmetries \cite{Li2008}.

It is also instructive to discuss the laser power/intensity scaling of the beam combining mechanism discussed above. We already showed that the pulses with slightly overcritical power are able to combine, even though the theoretical analysis suggests otherwise. Here, we seek the parameter regime where pulses merge despite being strongly overcritical and $a_0 >1$. In such a regime, the theory \cite{Berge1998} does not formally apply due to $|a|^2\ll 1$ approximation used in the derivation of NSE. At the same time, some early works\cite{Askaryan1994,Askaryan1997} suggest that $\lambda=1\rm um$, $a_0=5$, $w=9 \rm um$ laser pulse interacting with $n_e/n_{\rm cr}\approx 0.5$ uniform plasma slab ($P>>P_{\rm cr,2D}$) leads to pulse breakup into multiple filaments, which eventually merge into a single tightly focused filament with high energy conversion efficiency. Thus, there might be a regime where $P\gg P_{\rm cr,2D}$ pulses may combine as well.

\begin{figure}
    \centering
    \includegraphics[width=\linewidth]{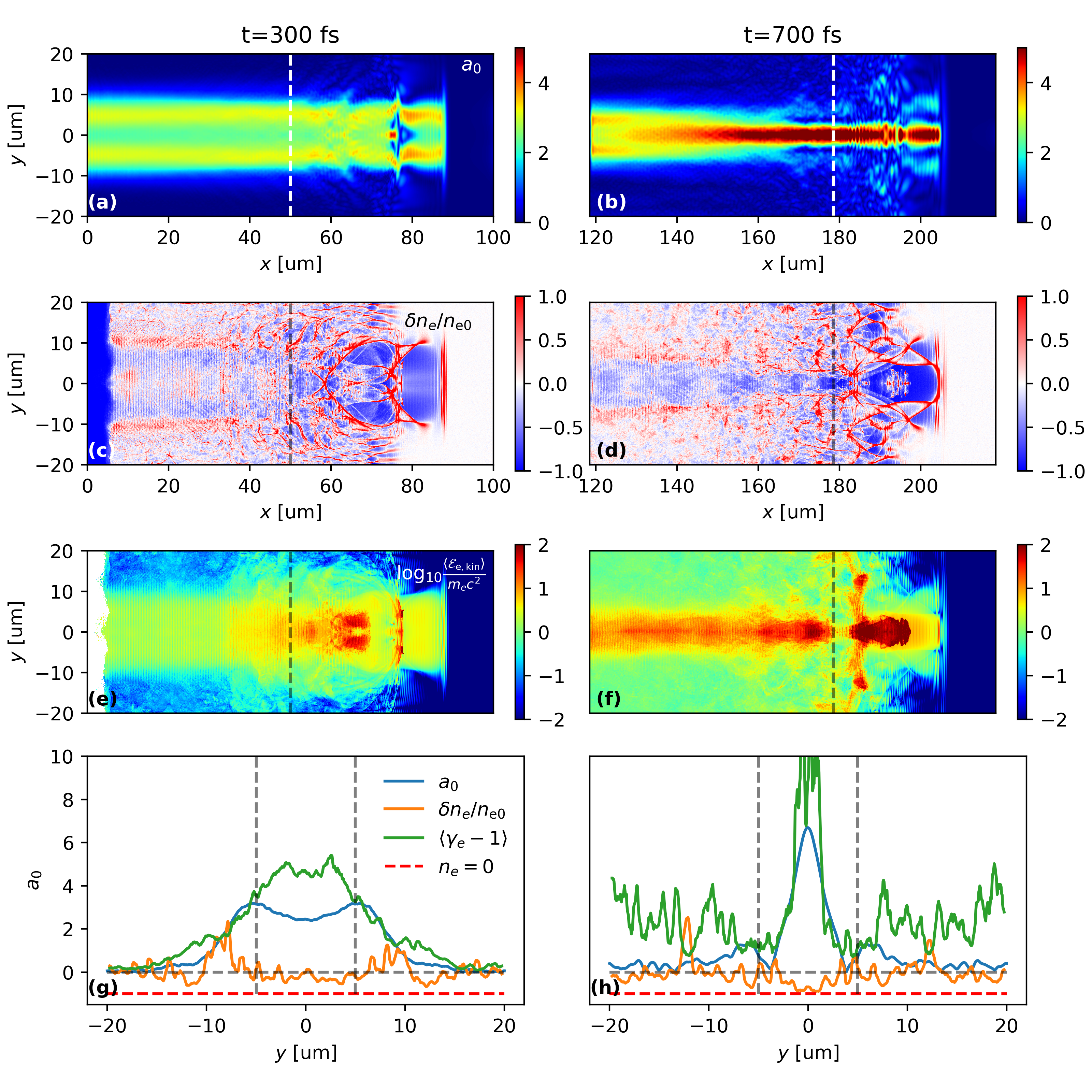}
    \caption{Snapshot of a simulation with $a_0=3$, $w=\delta=5 \rm um$, $\tau = 300 \rm fs$ ($P/P_{\rm cr,2D}  \approx 9$) at t=300 and 700 fs. Laser envelopes, density perturbations, the electron mean energization, and 1D cuts at $x=50 \rm um$ and $178 \rm um$ are shown. A beam merger is observed, even though both pulses are strongly overcritical.}
    \label{fig:a3_300fs}
\end{figure}

First, we reproduced results from Refs.\cite{Askaryan1994,Askaryan1997}, confirming the feasibility of the laser pulse filaments to recombine into a single tightly focused filament in a near-critical density plasma. As the next step, we considered $n_e/n_{\rm cr}=0.015$ plasma and two $a_0=3$, $\lambda=0.8 \rm um$, $w=\delta =5\rm um$, $\tau =300 \rm fs$ pulses. Figure~\ref{fig:a3_300fs} depicts the states of plasma profiles before and after beam merger. These pulses possess $P\approx 9 P_{\rm cr,2D}$ each, and tend to focus on their own early on in the run, as one may see from the leading edges of the pulse envelope figure at t=300 fs (Figure~\ref{fig:a3_300fs}a). Due to large laser fields ($a\geq 3$), plasma perturbations are strongly nonlinear, as one may see in density perturbation panels (Figure~\ref{fig:a3_300fs}c,d), with plasma bubble structure observed at the pulse leading edges. Electron heating is rapid and reaches ultrarelativistic energies of $ \gamma_{\rm max} \sim 10^2$ within the bubble structure and mean box-averaged electron energy being $\langle \gamma_e -1 \rangle \approx 1.7 $ and $5.0$ at $t=300$ and 700 fs, respectively (Figure~\ref{fig:a3_300fs}e,f). Strong density cavitation and hot electron structure around the $y=0$ axis lead to pulse combination, in qualitative agreement with the simulation with $a_0<1$ presented earlier in the paper, but with different electron energization dynamics strongly deviating from the theoretical model used in Figure \ref{fig:PICmerging_mechanism}. The details behind strong electron energization exceeding $m_ec^2 a_0$ estimate of electron energization in the laser field of two parallel beams are to be studied separately.

We thus conclude that the laser beam merger in tenuous plasmas is feasible for $P\sim P_{\rm cr}$ for laser beam separations given by Eqn.~\ref{eqn:mergercond3D} and, in some cases, even for $P\gg P_{\rm cr}$.

\section{Discussion}
\label{sec:diss}

In this paper, we addressed the question of the merger of parallel laser beams propagating in tenuous plasma. We revised the theoretical threshold of beam combination and verified it using 2D NSE simulations. We highlighted the physics of the beam combination on the basis of refractive index perturbations, and demonstrated the difference between NSE and full PIC physics, illustrating how two beam system with $a_0 \sim 1$ leads to a more complex behavior than the NSE model predicts. We showed that nonlinear density perturbations are the main factor differentiating NSE and PIC behavior, with density perturbations in PIC acting to merge slightly overcritical pulses, whereas density perturbations in NSE were small yet counteracting beam amalgamation. Three-dimensional PIC simulations confirm the possibility of generalizing our primarily two-dimensional results for real-world applications.

While the results in the manuscript were obtained for the short pulses of subpicosecond duration, the main conclusion about the possibility of the parallel beam combination in plasmas may be extended to long pulses as well. Indeed, even though the mechanism of transverse beam profile modulations could differ, be it relativistic, ponderomotive, or thermal focusing (see, e.g., Ref.~\cite{Li2019}), transverse dynamics of the beams would still be controlled by an equation similar to Eqn.~\ref{eqn:nsepdmtv1} (see  Eqn.~21a in Ref.~\cite{Li2019}), and one could in principle conduct a calculation similar to ours, finding the balance between diffraction and self-focusing to create a self-merging system of long beams. Thus, we believe our results are of possible interest for multi-beam facilities like NIF and OMEGA, where beam combining experiments utilizing multiple crossing beams were successfully conducted \cite{Kirkwood2018,Kirkwood2022}. For such systems, the aforementioned feature of the suppressed pulse power losses due to the effective decrease of the peak laser field up to the moment of beam combination would be especially beneficial.

Although the phase shift between the laser beams does not explicitly appear anywhere in the manuscript, it is an important parameter for the actual implementation of the beam combiner. Indeed, as it was shown for Kerr medium in Ref.~\cite{Ishaaya2007}, once the absolute value of a phase shift exceeds $\pi/4$, laser beams no longer merge and may even repel. One may think about the phase shift appearing in the $|a_1+a_2|^2$ term and once it is chosen in a way to reduce the magnitude of $|a_1+a_2|^2$ term, the refractive index in between the two beams becomes smaller, impeding beam merger. We reproduced such results with our auxiliary NSE simulations and with a low-resolution 2D PIC scan (20 grid cells/micron, 20 particles per cell), although PIC simulations suggest a smaller phase shift threshold for beam merger, $|\Delta \phi| \leq \pi/12$. Thus, the discussed beam merger mechanism may be thought of as a mode selector mechanism, combining pulses of identical phases and repelling pulses with a significant phase shift.

From the experimental perspective, it is of interest to address the question of the beam combination of oblique pulses via the mutual-focusing-like instability. For the obliquely overlapping beams, the crossing time may be estimated as $t_{\rm cross}=w/ c \sin{\theta}$ ($w$ being beam width and $\theta$ being crossing angle), which yields the ratio of combination time $\tau_{\rm MF} \sim \tau_{\rm SF} \sim {2w_a}/(ca_0 \sqrt{n_e/n_{\rm cr}})$\cite{Kelley1965,Fibich} to crossing time for small $\theta$:
${\tau_{\rm MF}}/{t_{\rm cross}} \sim {\theta \omega}/{\sqrt{2} a_0 \omega_{\rm pe}},$ which stays around 1 for $\theta = 1^\circ - 5^\circ$, $a_0 \sim 0.1$ and $\omega_{\rm pe}/\omega \sim 0.2$. Since the actual merger takes a few $\tau_{\rm MF}$'s, beam combination requires $\tau_{\rm MF}/t_{\rm cross} \ll 1$, imposing a severe restriction on crossing angle for beam combination. This could be potentially overcome by using a plasma channel with concave density distribution, acting as a defocusing lens. For the density distribution of $n_e = n_{e0}(1+y^2/l^2)$, with $l$ being channel width, one may estimate the length of the structure scattering laser rays from $\pm \theta$ to $0^{\circ}$ as $L_{\rm struct} \approx  l/(n_e/n_{\rm cr})^{1/2}$. Auxiliary NSE simulations of beam dynamics in the transverse plane (2D+1T) reveal that for small inclination angle ($\theta \lessapprox 2^\circ$) between the beams and for the beam pair's parameters specified above, we observe beam collapse at the center of symmetry during the beam crossing time. For larger angles, $\theta > 10^\circ$, the overlap is not long enough for the beam merger. Thus, the limitation of the small inclination angle may be overcome by crossing beams at $\theta \sim 1^\circ$ and/or using a defocusing lens-like structure.

It is natural to check whether we could apply the beam combination mechanism to $N>2$ beams. 3D PIC simulation showed that we do see a combination of three pulses separated by $2w$ each. Figure~\ref{fig:PIC3D}b demonstrated how three pulses of the total power of 1.5 TW were combined into a single beam with similar energy losses as in the two-pulse case. Auxiliary NSE simulations of beam dynamics in the transverse plane show that hexagonal structures of six beams with beam separation of $2w$, $a_0=0.17$, and $n_e/n_{\rm cr}=0.032$ combine into one around the hexagon center. If we consider a laser pulse array with uniformly distributed pulses, we may expect that beams on the edges will combine first - beams within the center will experience net zero mutual focusing, which could lead to the beam collapsing further away from the center of mass of the laser array. A more detailed analysis of the parallel laser beam array dynamics is needed to optimize beam array combination, which is beyond the scope of the present manuscript.

The results obtained in this paper may be of interest to a broad laser-plasma interaction community, including plasma-based laser amplification, plasma optics, and inertial fusion energy.

\section*{Acknowledgements}

This work was supported by the U.S. Department of Energy under contract number DE-AC02-09CH11466. The United States Government retains a non-exclusive, paid-up, irrevocable, world-wide license to publish or reproduce the published form of this manuscript, or allow others to do so, for United States Government purposes. This work was supported in part by NNSA DE-SC0021248. K.V.L. was partially supported by the Laboratory Directed Research and Development (LDRD) Program of Princeton Plasma Physics Laboratory. The EPOCH code was developed as part of the UK EPSRC funded projects EP$/$G054940$/$1. The simulations presented in this article were performed on computational resources managed and supported by Princeton Research Computing at Princeton University.

\end{document}